\documentclass[ reprint, amsmath,amssymb, aps, prb,superscriptaddress]{revtex4-1}
\usepackage{graphicx}% Include figure files
\usepackage{dcolumn}% Align table columns on decimal point
\usepackage{bm}% bold math
\usepackage{hyperref}% add hypertext capabilities
\usepackage{physics}
\usepackage{tabularx}
\usepackage{amsmath}
\usepackage{amssymb} %maths
\usepackage[utf8]{inputenc} %useful to type directly diacritic characters
\usepackage[T1]{fontenc}
\usepackage{xcolor}
\newcommand {\red}[1] {{\textcolor{black}{#1}}}

\begin{document}

\title{ Electron-phonon coupling and electronic thermoelectric properties of n-type {PbTe} driven near the soft-mode phase transition via lattice expansion
}

\author{Jiang Cao}
\email{jiang.cao@njust.edu.cn}
\affiliation{School of Electronic and Optical Engineering, Nanjing University of Science and Technology, Nanjing 210094, China}
\affiliation{Tyndall National Institute, Dyke Parade, Cork T12R5CP, Ireland}
\author{{\DJ}or{\dj}e Dangi\'c}
%\email{djordje.dangic@tyndall.ie}
\affiliation{Tyndall National Institute, Dyke Parade, Cork T12R5CP, Ireland}
\affiliation{Department of Physics, University College Cork, College Road, Cork T12K8AF, Ireland}
\author{Jos\'e  D. Querales-Flores}
%\email{jose.querales@tyndall.ie}
\affiliation{Tyndall National Institute, Dyke Parade, Cork T12R5CP, Ireland}
%\author{\blue{Olle Hellman}}
%\affiliation{\blue{Department of Physics, Chemistry and Biology (IFM), Link{\"o}ping University, SE-581 83, Link{\"o}ping, Sweden}}
\author{Stephen Fahy}
%\email{s.fahy@ucc.ie}
\affiliation{Tyndall National Institute, Dyke Parade, Cork T12R5CP, Ireland}
\affiliation{Department of Physics, University College Cork, College Road, Cork T12K8AF, Ireland}
\author{Ivana Savi\'c}
\email{ivana.savic@tyndall.ie}
\affiliation{Tyndall National Institute, Dyke Parade, Cork T12R5CP, Ireland}

\date{\today}

\begin{abstract}
  IV-VI materials are some of the most efficient bulk thermoelectric materials due to their proximity to soft-mode phase transitions,  which leads to low lattice thermal conductivity. \red{It has been shown that the lattice thermal conductivity of PbTe can be considerably reduced by bringing PbTe closer to the phase transition e.g. via lattice expansion.} However, the effect of soft phonon modes on
%their
\red{the} electronic thermoelectric properties of \red{such system} remains \red{unknown.}
%poorly understood.
Using first principles calculations, we show that the soft zone center transverse optical phonons do not deteriorate the electronic thermoelectric properties of PbTe driven closer to the phase transition via lattice expansion due to external stress, \red{and thus enhance the thermoelectric figure of merit.}
\red{We find that the optical deformation potentials change very weakly as the proximity to the phase transition increases, but the population and scattering phase space of soft phonon modes increase. Nevertheless, scattering between electronic states near the band edge and soft optical phonons remains relatively weak even very near the phase transition.}
%as a result of symmetry restrictions. 
%In contrast, phonon softening considerably reduces the lattice thermal conductivity of
%PbTe with the lattice constants $\sim 1.5$\% larger than that at equilibrium and doubles the thermoelectric figure of merit.
%Our results \red{thus} indicate that the thermoelectric figure of merit  of materials with soft phonon modes that interact strongly with other phonons, but weakly with electronic states relevant for transport, could be significantly increased by bringing these materials closer to the soft-mode phase transition.
\end{abstract} 

\maketitle

\section{Introduction}

The metric by which the performance of a thermoelectric material is measured is the thermoelectric figure of merit, $ZT=(\sigma S^2 T)/(\kappa_e + \kappa_L)$, where $S$ is the Seebeck coefficient, $\sigma$ is the electrical conductivity, $T$ is the temperature, and $\kappa_e$ and $\kappa_L$  are  the electronic and lattice contribution to thermal conductivity, respectively~\cite{snyder_2008,he_2017,zhou_2018}. Achieving
high $ZT$ values requires a high power factor (PF=$S^2\sigma$) and low $\kappa_e$ and $\kappa_L$. The electronic transport properties $S$, $\sigma$ and  $\kappa_e$ are correlated, while $\kappa_L$  is to some degree independent from the electronic properties in semiconductors.

Reducing $\kappa_L$ without changing electronic transport properties in a material is a frequently used approach for improving $ZT$, known as the “phonon glass-electron crystal”~\cite{Beekman2015,Takabatake2014}. It has been recently recognized that the proximity to soft-mode phase transitions is one of the key ingredients in suppressing the lattice thermal conductivity of IV-VI thermoelectric materials~\cite{delaire_2011,Shiga2012,li_2014,li_2015,zhao_2014,Romero2015,Murphy2016}. Bringing SnSe near its soft-mode phase transition via temperature has been shown to enhance anharmonicity and reduce $\kappa_L$~\cite{zhao_2014,Aseginolaza2019,Chang2019,HONG2019}. Similar effects have also been predicted for PbTe driven to a different type of
 the soft-mode phase transition via strain or alloying~\cite{Romero2015,Murphy2016,Skelton2014,Murphy2017}. An important open question is whether
soft phonon modes can lead to strong electron-phonon (e-ph) scattering,
and actually
reduce the PF, thus competing with the benefit of lowering $\kappa_L$ for improved thermoelectric performance.

We have recently shown using first principles calculations that soft phonon modes in PbTe interact weakly with the electronic states relevant for thermoelectric transport~\cite{Cao2018,DSouza2020}. 
Such weak e-ph scattering was partially attributed to the symmetry forbidden 
 conduction band intra-valley scattering (by the $\Gamma$-point phonons).  
However, this symmetry restriction at L (the band minimum) becomes insufficient in explaining the strength of e-ph coupling  as the material approaches the soft-mode phase transition. In this case, the soft mode frequency becomes nearly zero and the soft modes exhibit almost an acoustic-like dispersion relation near the Brillouin zone center. As the proximity to the soft-mode phase transition increases, scattering due to soft optical phonon modes becomes increasingly like that due to acoustic phonons, which is also forbidden at the band minimum L but nevertheless relatively strong for carriers near L, which are active in thermoelectric transport in PbTe.
  Moreover, the optical deformation potentials  are much larger than those of acoustic phonons in equilibrium PbTe~\cite{Cao2018}, which could make electron-soft mode scattering significantly stronger closer to the phase transition if the optical deformation potentials are not significantly reduced.
As PbTe approaches the phase transition, the decreasing TO phonon frequency will significantly increase the TO phonon population and the e-ph scattering phase space, thus increasing the e-ph scattering due to the soft modes.
For example, recent first principles calculations in SrTiO$_3$ have revealed that soft phonon modes become the dominant e-ph scattering channel near the Curie temperature~\cite{Zhou2018}. Whether soft modes become the dominant electronic scattering channel in a particular material depends on the exact magnitude of deformation potentials close to the phase transition, which cannot be known apriori from the symmetry analysis alone and calls for first-principles calculations.

In this paper, we show from first principles calculations that driving n-type PbTe close to its soft-mode phase transition via lattice expansion due to external stress does not degrade the electronic transport properties. 
We find that optical deformation potentials do not change much with strain. Since the soft mode populations and scattering phase space increase as PbTe is driven closer to the phase transition, soft TO modes interact more strongly with the electronic states determining thermoelectric transport. However, soft TO modes remain a relatively weak e-ph scattering channel compared to long-range polar longitudinal optical (LO) phonon modes. Consequently, increasing the proximity to the soft TO mode phase transition via lattice expansion does not have a detrimental effect on the electronic thermoelectric properties of PbTe.
 This result, together with the earlier predictions of a significant $\kappa_L$ decrease in PbTe driven closer to the phase transitions (roughly by a factor of 2)~\cite{Romero2015}, indicates a potential considerable increase of $ZT$ near the phase transition.

\section{Method}
\label{sec:method}

We employed the thermoelectric transport model developed in our previous work and tested for PbTe against many experiments~\cite{Cao2018,Cao2019,DSouza2020}. We calculated all the parameters of the model from first principles, thus relying on no free parameters. The accuracy of our transport model in calculating the e-ph relaxation times using the density functional theory (DFT) electronic band structure is similar to that of the state-of-the-art EPW code~\cite{Ponce2016}, as shown in Sec.~\ref{ssec:results} and Refs.~\cite{Cao2018,DSouza2020}. The results of our model are directly comparable with those of the EPW code when we use parametrized electronic and phonon band structures obtained with DFT and density functional theory (DFPT)~\cite{Gonze1997,Baroni2001}, respectively, and the Quantum Espresso package~\cite{QE-2009,QE-2017}.
However, unlike the current version of the EPW code, our transport model allows using a more accurate electronic band structure than that computed with DFT and accounting for its temperature dependence, as described in the following.

\red{In order to get the electronic band structure with an accurate bandgap and the correct character of conduction and valence band edge states, we used} the screened Heyd-Scuseria-Ernzerhof (HSE03) hybrid functional including SOC~\cite{HSE2003,HSE2004} and the VASP package~\cite{KRESSE1996}.  The basis set for the one-electron wave functions was constructed with the projector augmented wave  method~\cite{Kresse1999}.
\red{In PbTe, the direct band gap is located at the L points in the Brillouin zone and only the electronic states in the vicinity of the L points contribute to electronic transport~\cite{Cao2018}.}
The HSE03 lowest conduction band was fitted to the two-band Kane model near the L points,
using the parallel and perpendicular effective masses ($m^*_{\perp,\parallel}$) and the non-parabolic parameter set to the inverse of the bandgap ($1/E_g$)~\cite{Cao2018}.

\red{In our electronic transport model, we also accounted for the temperature dependence of the bandgap due to e-ph interactions using the non-adiabatic Allen-Heine-Cardona (AHC) formalism~\cite{Allen1976,Allen1981,Allen1983} and DFPT~\cite{Gonze1997,Baroni2001}, as implemented in ABINIT~\cite{Ponce2015}}. We calculated $\partial E_g/ \partial T=2.98\times10^{-4}$~eV/K\cite{querales-flores_2019}. The effective masses are taken to depend on temperature as $m^*_{\perp,\parallel}(T)/m^*_{\perp,\parallel}({\rm 0K})=E_g(T)/E_g({\rm 0K})$ according to the two-band Kane model~\cite{Dresselhaus2008a}. We note that the thermal expansion contribution to $\partial E_g/ \partial T$ is excluded in this study since we explicitly investigate the effect of lattice expansion on thermoelectric transport. 
  
To calculate phonon band structures and e-ph matrix elements, we
%first computed e-ph matrix elements within
adopted the DFPT framework as implemented in the QUANTUM ESPRESSO package~\cite{QE-2009,QE-2017}, using the LDA excluding SOC with the HGH norm-conserving pseudopotentials. In our previous work, we verified that the LDA without SOC gives very similar e-ph matrix elements compared with the HSE03 with SOC~\cite{murphy_2018}. For the LDA with SOC, the e-ph matrix elements are ill-defined, because the conduction band minimum and valence band maximum states are interchanged and mix heavily, forming an incorrect "negative bandgap"~\cite{murphy_2018}. In our DFPT calculations of e-ph matrix elements, we used 12$\times$12$\times$12 and 6$\times$6$\times$6 reciprocal space grids for electronic and phonon states, respectively, and the cutoff energy was set to 70 Ry. E-ph matrix elements were then interpolated on finer 40$\times$40$\times$40 $\bm{k}$ and $\bm{q}$-grids using the Wannier functions method~\cite{Giustino2007} and the EPW code~\cite{Ponce2016}. Fourteen Wannier orbitals were constructed from Bloch states on a 12$\times$12$\times$12 $\bm{k}$-point grid using the Wannier90 code~\cite{MOSTOFI2014}.

In PbTe, scattering at the L-points caused by the X-point phonons is forbidden by symmetry~\cite{Cao2018}. 
	Moreover, X-point phonon frequencies remain non-zero throughout the range of lattice and temperature parameters of our study.
	As a result, only phonons near the zone center contribute to electronic scattering processes in n-type PbTe in our model, which is verified against rigorous EPW calculations (see Sec.~\ref{ssec:results}). 
	We parametrized nonpolar and polar Fr\"ohlich e-ph scattering mechanisms using the approach developed in our previous work~\cite{murphy_2018,Cao2018}. Acoustic and optical deformation potentials were calculated by fitting the corresponding e-ph matrix elements with the deformation potential Hamiltonian in the long-wavelength limit along five high-symmetry directions of $\bm{q}$ in a cubic crystal~\cite{murphy_2018,Fahy2008,Herring1956}. Ionized-impurity scattering induced by dopant atoms was accounted for using the Brooks-Herring model~\cite{Ridley1999} with the Thomas-Fermi screening. Elastic and dielectric constants for the polar Fr\"ohlich perturbation Hamiltonian were obtained from DFPT.

Electronic transport quantities were calculated using the Boltzmann transport theory within the transport (momentum) relaxation time approximation as
\begin{equation}
\label{eq:transport}
\begin{cases}
\sigma       & = L_0 , \\
S                & = - L_1 / (eT L_0) , \\
\kappa_0   & = L_2 / (e^2 T) ,
\end{cases}
\end{equation}
where $\sigma$ is the electrical conductivity tensor, $S$ is the Seebeck coefficient tensor, $\kappa_0$ is the thermal conductivity tensor defined when the electric field across the material is zero, and $e$ is the electron charge value. The transport kernel functions for the conduction band are defined by
\begin{equation}
\label{eq:kernal}
L_\alpha = \int_{\rm BZ} \frac{ e^2 d \bm{k}}{4 \pi^3} 
\left( - \frac{\partial f}{\partial E} \right) \tau_{\bm{k}{,\rm tot}} 
\bar{v}^2_{\bm{k}} \left( E_{\bm{k}} - E_F \right)^\alpha,	
\end{equation}
where $E_{\bm{k}}$ is the energy of the conduction band state with the crystal momentum  $\bm{k}$, $\bar{v}_{\bm{k}}= [\sum_i (v^i_{\bm{k}})^2/3]^{1/2}$ is 
the group velocity averaged over directions, $i$ denotes the Cartesian directions, $f$ is the Fermi-Dirac distribution function, $E_F$ is the Fermi level, and $\tau_{\bm{k}{,\rm tot}}$ is the total transport relaxation time of all scattering channels~\cite{Cao2018}.

\section{Results and discussion}
\label{ssec:results}

\subsection{Electronic band structure}

\begin{figure}[t]
	\centering
	\includegraphics[width=8cm]{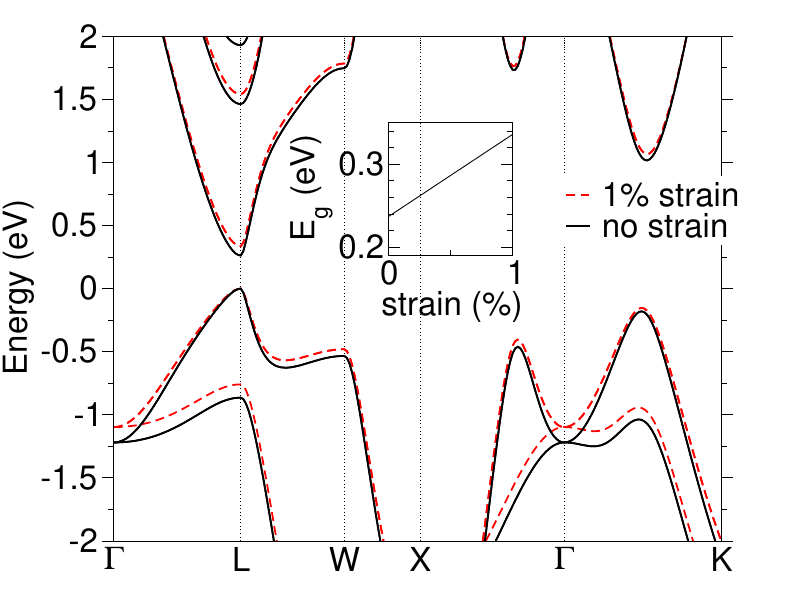}
	\caption{ Electronic band structure at
          0~K for PbTe
          with the equilibrium lattice constant (solid black lines) and
          the $1$\% larger lattice constant
          (dashed red lines). The inset shows direct bandgap versus strain. }
	\label{fig:ebands}
\end{figure}

PbTe has
the conduction band minima (CBM) and the valence band maxima (VBM)
located at the four L points in the BZ. In Fig.~\ref{fig:ebands}, we plot the electronic band structure calculated using the HSE03 hybrid functional
for PbTe with the equilibrium and $1$\% expanded lattice constants ($a_{\rm latt}=a_0$ and $a_{\rm latt}=1.01 a_0$, respectively).
Most importantly, strain has a fairly weak effect on the
electronic band structure close to the Fermi energy. The direct bandgap at the L points ($E_g$) is increased from 0.238~eV to 0.336~eV by
applying 1\% of strain, as shown in the inset
of Fig.~\ref{fig:ebands}. The conduction and valence band effective masses also increase with the bandgap.
We can accurately fit the HSE03 conduction band within
0.5~eV from the CBM
using the Kane model if we use $1/E_g$ as the non-parabolic parameter of the Kane model, and the effective masses calculated using the LDA without SOC given in Table~\ref{tab:parameters}.

\begin{table}[t!]
\centering
\begin{tabular}{c c c}
	\hline 
	\hline
	$\eta$ & 0 & 1\% \\ 
	\hline 
	$a_{\rm latt}$ (\AA) & 6.348 & 6.411 \\
	$\Xi^{\rm ac}_d$ (eV) & 0.37 & 0.36 \\ 
	$\Xi^{\rm ac}_u$ (eV) & 7.03 & 6.48 \\ 
	$\Xi^{\rm opt}_d$ (eV) & 19.09 & 19.56 \\ 	
	$\Xi^{\rm opt}_u$ (eV) & -34.24 & -34.79 \\ 		
	$\omega_\Gamma^{\rm LO}$ (THz) & 3.2 & 3.1 \\
	$\omega_\Gamma^{\rm TO}$ (THz) & 0.95 & 0.15 \\
	$q_0$ (nm$^{-1}$)  &  1.61 &  0.254 \\
	$\epsilon_\infty$ & 34.85 & 33.59 \\
	$\epsilon_s$ & 356.8 & 14346 \\
	$c_{11}$ (GPa) & 136.4 & 120.6 \\
	$c_{12}$ (GPa) & 3.8 & 4.5 \\
	$c_{44}$ (GPa) & 17.1 & 17.0 \\
	$E_g^{\rm HSE}$ (eV) & 0.238 & 0.336 \\	
	$m^*_\parallel/m_0$ & 0.216 & 0.241  \\
	$m^*_\perp/m_0$ & 0.037 & 0.043 \\
	\hline 
	\hline
\end{tabular} 
\caption{Parameters used in the calculation of electronic transport quantities and computed from first principles for n-type PbTe with the equilibrium
  lattice constant (no strain, $\eta=0$) and  the lattice constant expanded by 1\%
  ($\eta=1 \%$ of strain):
  lattice constant in the 
local density approximation
  ($a_{\rm latt}$), acoustic deformation potentials ($\Xi^{\rm ac}_d$ and $\Xi^{\rm ac}_u$), optical deformation potentials ($\Xi^{\rm opt}_d$  and $\Xi^{\rm opt}_u$), optical phonon frequencies ($\omega_\Gamma^{\rm LO}$ and $\omega_\Gamma^{\rm TO}$), TO phonon dispersion parameter  ($q_0$),  high-frequency and static dielectric constants ($\epsilon_\infty$ and $\epsilon_s$), elastic constants ($c_{11}$, $c_{12}$ and $c_{44}$), direct bandgap from the
  hybrid functional calculations ($E_g^{\rm HSE}$), and parallel and perpendicular effective masses at the conduction band minimum
  ($m^*_\parallel$ and $m^*_\perp$).
}
\label{tab:parameters}
\end{table}

\subsection{Transverse optical mode softening}

\begin{figure}[t]
	\centering
	\includegraphics[width=8cm]{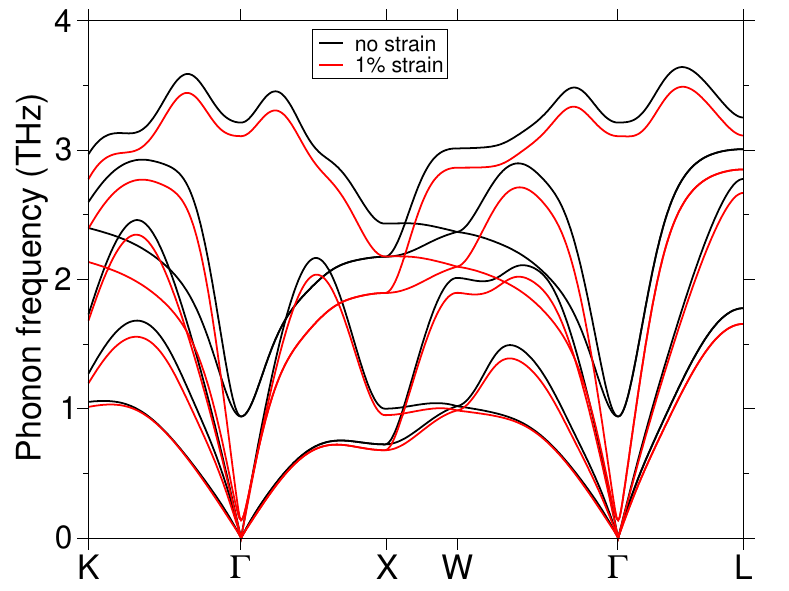}
	\caption{ Phonon dispersion
		for PbTe with the equilibrium lattice constant (black lines) and
		the $1$\% larger lattice constant (red lines),
                %		Solid lines represent the results computed using the temperature dependent effective potentials (TDEP) method at 50 K, while dotted lines are
                obtained using density functional perturbation theory (DFPT). 
	}          
	\label{fig:phbands}
\end{figure}

\begin{figure}[t]
	\centering
	\includegraphics[width=8cm]{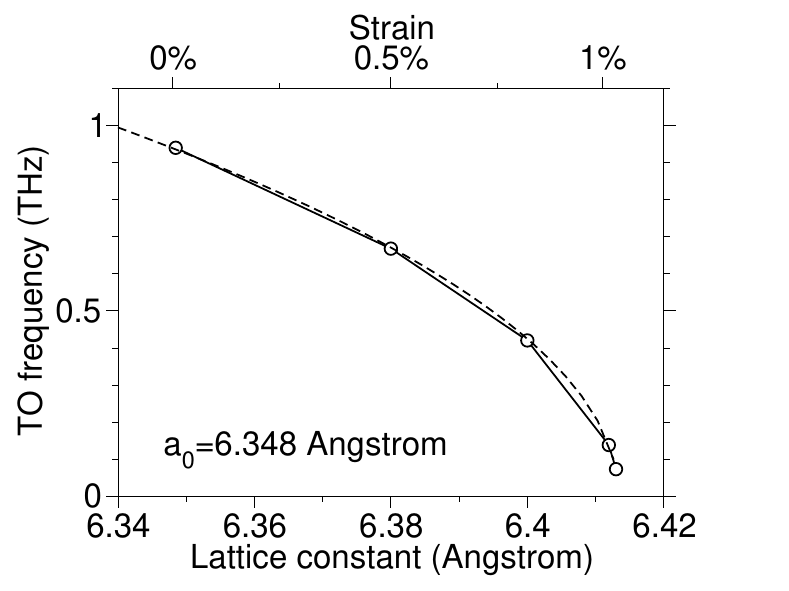}
	\caption{  
		Frequency of transverse optical (TO) modes at the zone center
		for PbTe versus lattice constant and strain, calculated using density functional perturbation theory (DFPT).
                %, black circles)  and the temperature-dependent effective potential (TDEP) method at 50~K (red squares).
                The dashed black line represents a fit to the DFPT results. 
            }
	\label{fig:TOfreq}
\end{figure}

PbTe crystallizes in the rocksalt structure up to the melting temperature of $\sim 1200$~K~\cite{ravich2013semiconducting}. Nevertheless, it is energetically close to the phase transition to the rhombohedral phase, characterized by the internal atomic displacement of Te, which corresponds to a frozen-in atomic motion of the zone center TO mode along the [111] direction in the rocksalt structure. As a result, the TO phonon modes near the zone center in PbTe are very soft~\cite{Jiang2016,delaire_2011}, see the black solid lines in Fig.~\ref{fig:phbands}. If the lattice is expanded via external stress, the frequencies of the soft TO modes near the $\Gamma$ point further decrease (the red solid lines in Fig.~\ref{fig:phbands}), since the potential energy surface flattens as PbTe approaches the soft-mode phase transition.
To gain more insight into the
TO mode behavior near the soft-mode phase transition, Fig.~\ref{fig:TOfreq}
shows the TO mode frequency at the $\Gamma$ point, $\omega_{\rm TO}^\Gamma$,
as a function of the lattice constant and the corresponding strain.
%The
$\omega_{\rm TO}^\Gamma$
%computed from DFPT
can be fitted to
$\omega_{\rm TO}^\Gamma = \alpha \sqrt{a_c - a_{\rm latt}}$, with $\alpha=3.67246$~THz~\AA$^{-0.5}$  and $a_c=6.41338$~\AA, shown by the black dashed line in  Fig.~\ref{fig:TOfreq}.
LO and acoustic modes are much less sensitive to changes of the lattice constant compared to soft TO modes, Fig.~\ref{fig:phbands}.

\subsection{Electron-phonon interaction}

\begin{figure*}[t]
	\centering
	\includegraphics[width=6cm]{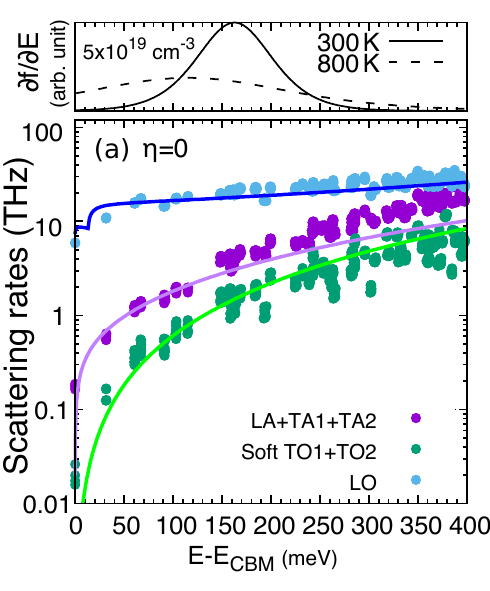}		
	\includegraphics[width=6cm]{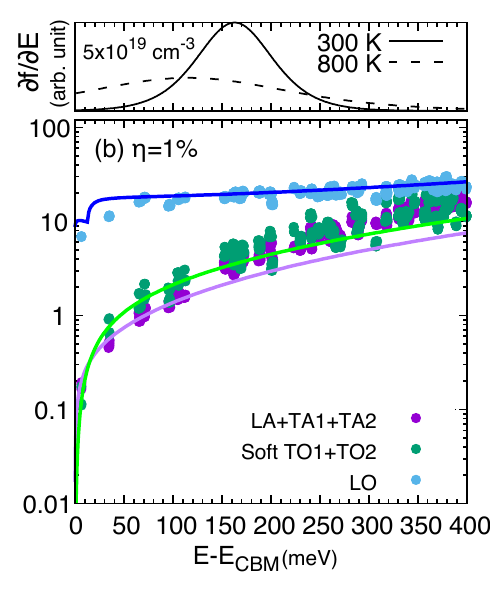}	
	\caption{ Electron-phonon scattering rates at 300~K resolved by phonon mode as a function of
		electronic energy for PbTe
		with (a) no applied strain ($\eta$=0) and (b) 1\% of strain ($\eta$=1\%).  The energy derivative of the Fermi distribution function
		for the doping concentration of 5$\times10^{19}$~cm$^{-3}$ is also plotted for 300~K and 800~K. The zero of the energy axis is the
		conduction band minimum.}
	\label{fig:scattering}
\end{figure*}

To accurately describe the e-ph interaction
in n-type PbTe driven close to the phase transition via lattice expansion, we calculated all the
relevant quantities~\cite{Cao2018} from first principles for
the equilibrium and 1\% expanded lattice constants, as reported in Table~\ref{tab:parameters}.
The elastic constants do not vary considerably with strain, reflecting the same trend seen for acoustic phonon branches, as well as the LO phonon frequency (Fig.~\ref{fig:phbands}). 
The only phonon quantities that change significantly while varying the amount of strain are the TO phonon frequency at $\Gamma$, $\omega_{\rm TO}^\Gamma$,
and the TO dispersion in the vicinity of $\Gamma$ (Fig.~\ref{fig:phbands}). 
We note that the TO dispersion near $\Gamma$ was modeled as $\omega_{\rm TO}^{\bm{q}}=\omega_{\rm TO}^\Gamma \sqrt{1+|\bm{q}|^2/q_0^2}$, which transforms smoothly from equilibrium to 1\% strain
  as $q_0$ decreases with $\omega_{\rm TO}^\Gamma$, see Table~\ref{tab:parameters}. 

The high frequency dielectric constant $\epsilon_\infty$ does not change much with lattice expansion since the electronic band structure is not substantially modified. However, the static dielectric constant $\epsilon_s$ diverges towards the phase transition as a result of the Lydanne-Sachs-Teller (LST) relation~\cite{LST}: $(\omega_{\rm LO}^\Gamma/\omega_{\rm TO}^\Gamma)^2 = \epsilon_s/\epsilon_\infty$.

\red{Since the electronic band structure and acoustic phonons of PbTe depend weakly on strain, the acoustic deformation potentials do not change much with strain, see Table~\ref{tab:parameters}. Surprisingly, despite the fact that $\omega_{\rm TO}$ varies significantly with strain, the optical deformation potentials are relatively insensitive to it. We emphasise that it is impossible to apriori know the variation of deformation potentials with strain without carrying out these first principles calculations.} In the following calculations, we will only update the values of $\omega_{\rm TO}^\Gamma$, $q_0$ and $\epsilon_s$ as a function of strain, as well as the parameters of the electronic band structure i.e. the bandgap and the effective masses. 

Next we compute the e-ph scattering rates of PbTe using our model, and compare them with those obtained using the EPW code, which makes no assumptions about the $\bm{k}$- and $\bm{q}$-point dependence of e-ph matrix elements and their changes with strain~\cite{Ponce2016}. 
In the EPW calculations, we used  40$\times$40$\times$40 $\bm{k}$- and $\bm{q}$-point grids, and the Gaussian broadening parameter was set to 40~meV.
Figure~\ref{fig:scattering} shows the mode-resolved e-ph scattering rates for n-type PbTe under different strain conditions at 300~K, calculated using our model and the EPW code.
We note that the model is in very good agreement with the EPW results for all the phonon modes and for different strain values. This finding highlights the accuracy of our model and further justifies our assumption of keeping the acoustic and optical deformation potentials, $\epsilon_\infty$ and the elastic constants constant. 

Scattering due to LO phonons is the strongest e-ph scattering mechanism in equilibrium PbTe, while scattering due to soft TO phonons is the weakest [Fig.~\ref{fig:scattering} (a)]. \red{In our previous work, we found that} the latter effect partly arises because the electron-TO mode matrix element is zero by symmetry for the conduction (and valence) band states at L and the zone center TO modes in the rocksalt structure~\cite{Cao2018}.
Similarly, the intervalley L-L scattering caused by the X-point phonons also vanishes~\cite{Cao2018}.

In spite of this symmetry restriction, it is apriori unclear whether the strength of electron-TO phonon scattering will increase as PbTe approaches the phase transition due to lattice expansion.
Extremely soft optical phonons behave almost like acoustic phonons, whose scattering at the zone center is also forbidden. Moreover, uniaxial optical deformation potentials are significantly larger than those of acoustic phonons (see Table~\ref{tab:parameters}) and scattering rates are proportional to their square. This suggests that scattering due to soft optical modes may become stronger near the phase transition.

We find that the strength of electron-TO phonon scattering indeed increases as PbTe approaches the phase transition, as shown in Fig.~\ref{fig:scattering}. In addition to the large values of optical deformation potentials, another reason for this effect is that
the phonon population and the vibration amplitude corresponding to soft TO modes increase. Furthermore, lowering the TO mode frequency near the zone center means that the energy and momentum conservation conditions for scattering become easier to satisfy, hence increasing the electron-TO phonon scattering phase space. The energy dependence of the TO scattering rates also changes significantly with lattice expansion, exhibiting  a more acoustic-like energy dependence close to the phase transition, see Fig.~\ref{fig:scattering}. As a result, the TO scattering rates near the phase transition are considerably higher for low-energy electrons close to the conduction band edge than those at equilibrium.

Longitudinal optical and acoustic phonon scattering are much less sensitive to strain compared to soft modes. \red{LO scattering is dominated by the polar Fr\"ohlich contribution. Even though the changes in optical deformation potentials increase the non-polar contribution to LO scattering, this increase is small compared to the polar contribution.} Most importantly, within the energy range relevant for electronic transport properties, polar LO phonon scattering is still stronger than that of acoustic and TO phonons even for 1\% strain. Due to this predominant role of polar LO phonon scattering, strain has a weak influence on the total e-ph scattering in PbTe.

\begin{figure}[t]
	\centering
	\includegraphics[width=8cm]{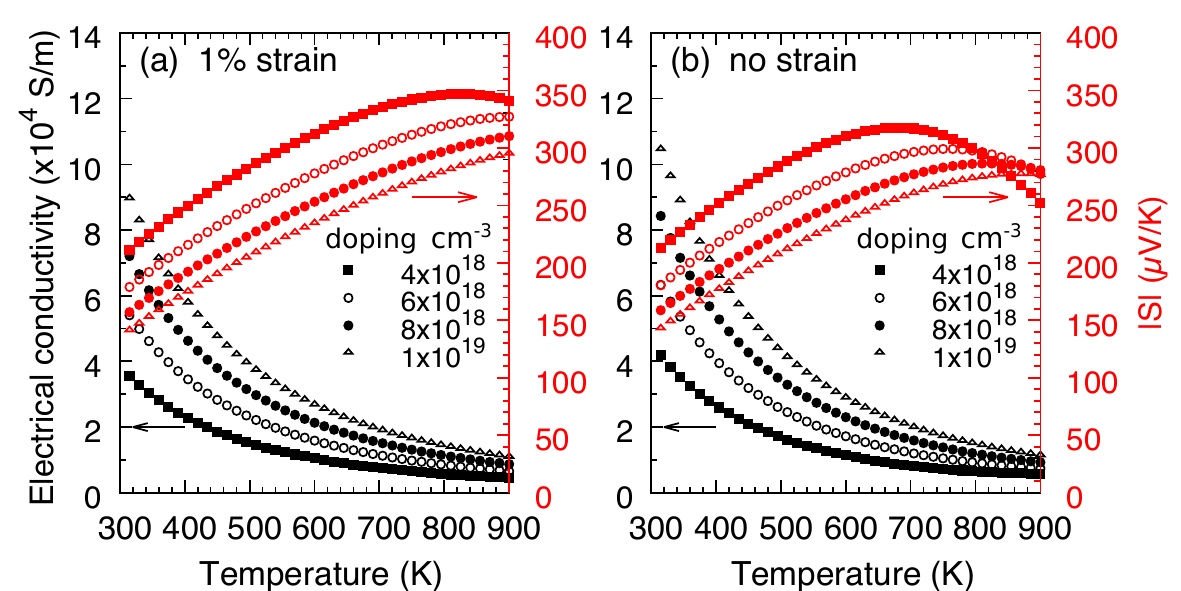}
	\caption{ Electrical conductivity and absolute Seebeck coefficient ($\abs{\rm S}$) as a function of temperature for PbTe: (a) with 1\% strain and (b) without strain, for the doping concentrations of 4$\times10^{18}$, 6$\times10^{18}$, 8$\times10^{18}$ and 1$\times10^{19}$~cm$^{-3}$. }
	\label{fig:transport-T}
\end{figure}

\subsection{  Electronic thermoelectric properties  }

To
elucidate the effect of the soft-mode phase transition on electronic transport in PbTe, we
illustrate the temperature dependence of the electrical conductivity and the Seebeck coefficient at different doping levels for PbTe with no strain and 
1\% of strain in Fig.~\ref{fig:transport-T}. Increasing the lattice constant results in a wider electronic bandgap (Fig.~\ref{fig:ebands}). This reduces the minority carrier concentration and the  bipolar effect that causes a \red{small} drop of the Seebeck coefficient at high temperatures and low doping concentrations\cite{Cao2019}. Therefore, strain leads to an improvement of the Seebeck coefficient
above 600~K. A larger bandgap due to strain also flattens the conduction band and leads to larger effective masses at the band edge, which explains the higher peak value of the Seebeck coefficient closer to the phase transition. On the other hand, due to larger effective masses, the electrical conductivity is
%inevitably
\red{slightly} reduced with lattice expansion. 
In addition, we note that the contribution of ionized impurity scattering to the total electron scattering rates is much smaller than that of phonons at room temperature and above in PbTe with no strain~\cite{Cao2018}. When TO modes soften, the static dielectric constant increases (see Table~\ref{tab:parameters}), thus leading to a reduction in ionized impurity scattering and a negligible effect on electronic transport.

\begin{figure*}
	\centering
	\includegraphics[width=15cm]{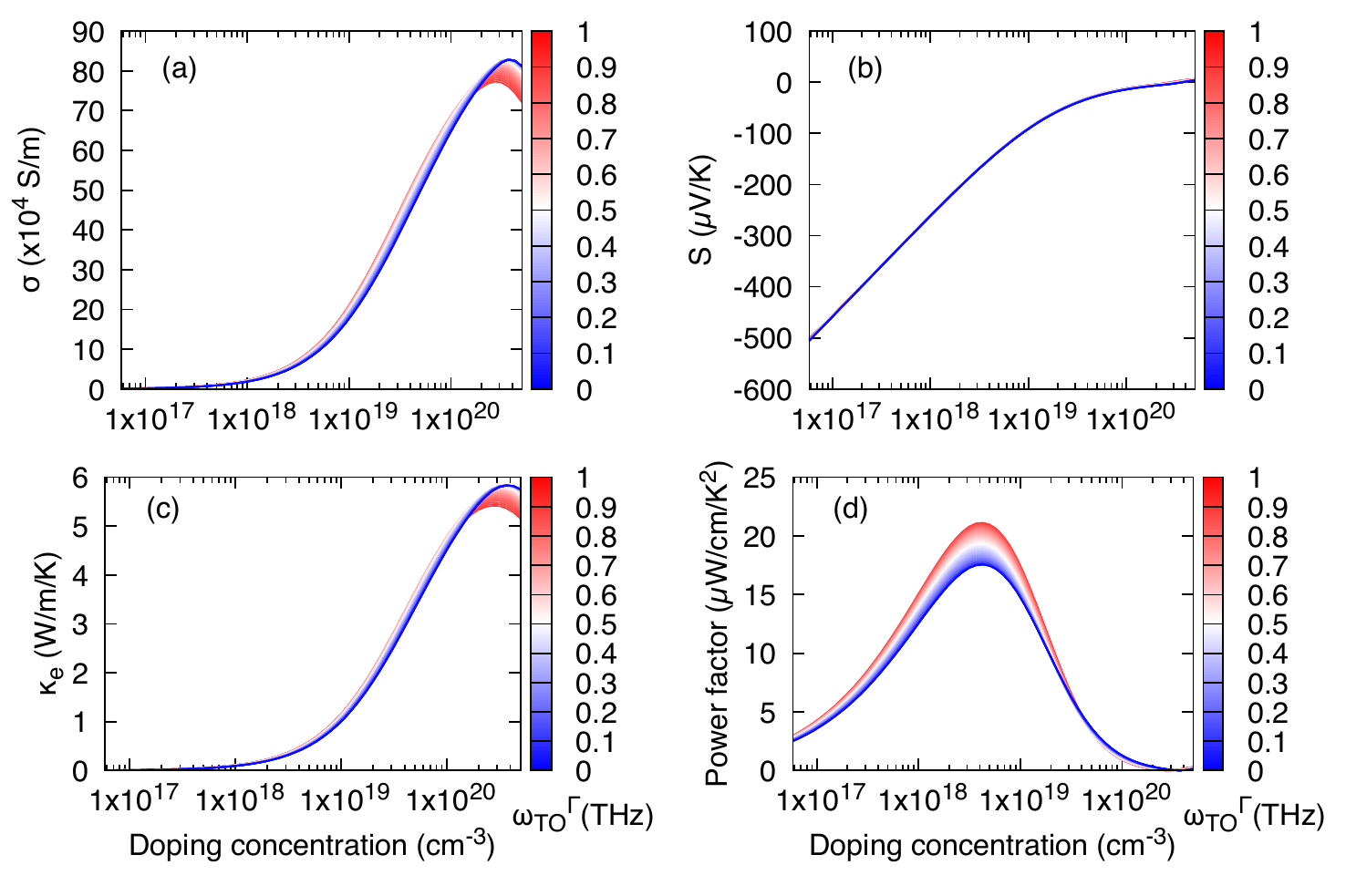}		
	\caption{ Thermoelectric transport properties as a function of doping concentration at 300~K for different transverse optical} (TO) phonon frequencies at the zone center obtained by applying strain: (a) electrical conductivity ($\sigma$), (b) Seebeck coefficient ($\abs{\rm S}$), (c) electronic thermal conductivity ($\kappa_e$),  and (d) power factor ($\sigma {\rm S}^2$).
	\label{fig:transport}
\end{figure*}

\begin{figure*}
	\centering
	\includegraphics[width=15cm]{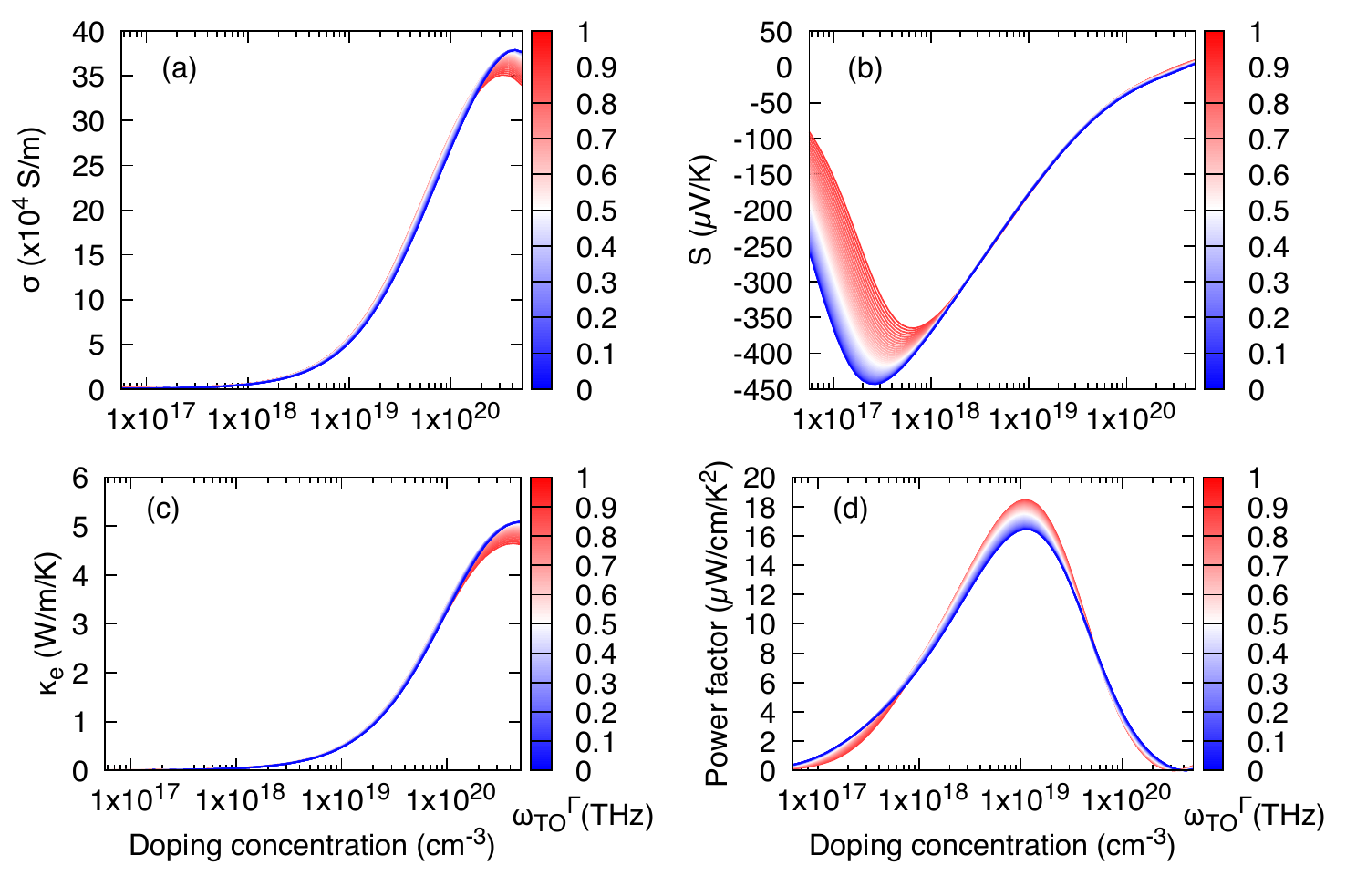}		
	\caption{ Thermoelectric transport properties as a function of doping concentration at 600~K for different transverse optical} (TO) phonon frequencies at the zone center obtained by applying strain: (a) electrical conductivity ($\sigma$), (b) Seebeck coefficient ($\abs{\rm S}$), (c) electronic thermal conductivity ($\kappa_e$),  and (d) power factor ($\sigma {\rm S}^2$).
	\label{fig:transport-600K}
\end{figure*}

To further
understand the role played by the closeness to the phase transition on the electronic thermoelectric transport properties of PbTe, in Fig.~\ref{fig:transport} and Fig.~\ref{fig:transport-600K} we 
plot all the relevant properties at 300 K and 600 K, respectively,
as a function of the doping concentration for different values of $\omega^\Gamma_{\rm TO}$, \red{which correspond to different values of strain indicated in Fig.~\ref{fig:TOfreq}.} The electrical conductivity and the electronic contribution to the thermal conductivity are
weakly influenced by the proximity to the phase transition for the doping concentrations lower than $1\times10^{20}$~cm$^{-3}$,
even though TO phonon scattering increases in magnitude.
The absolute value of the Seebeck coefficient
practically does not change with $\omega^\Gamma_{\rm TO}$ at 300~K, while it increases with decreasing $\omega^\Gamma_{\rm TO}$ for low doping concentrations at 600~K.
The latter is a consequence of the wider bandgap induced by
 strain, which reduces the bipolar contribution from the hole carriers.
The thermoelectric power factor
decreases slightly in the peak value closer to the phase transition, since the electrical conductivity is lower compared to equilibrium. Therefore, driving PbTe closer to the phase transition via lattice expansion does not appreciably degrade the electronic thermoelectric properties.

\subsection{Discussion}

In this work, we calculate the phonon frequencies and e-ph matrix elements that determine electronic thermoelectric transport using DFPT. These calculations show that although the TO frequency changes significantly due to strain, all other parameters characterizing electronic and phonon band structure and e-ph matrix elements that significantly affect electronic transport properties are fairly insensitive to strain (see Table~\ref{tab:parameters}). This means that even if the TO frequency is renormalized due to anharmonic effects and/or thermal expansion, our results corresponding to such renormalized frequency would still be similar to those reported in Fig.~\ref{fig:transport} and Fig.~\ref{fig:transport-600K}. Those figures clearly show that electronic thermoelectric properties are almost unaffected by the TO frequency
for doping concentrations where the power factor peaks.

\red{We also discuss the implications of driving n-type PbTe closer to the soft-mode phase transition on the thermoeletric figure of merit. Previous calculations of the lattice thermal conductivity $\kappa_L$ of PbTe show that $\kappa_L$ can be decreased roughly by a factor of $2$ if the lattice constant increases by 1\%~\cite{Romero2015}. The $\kappa_L$ decrease is a result of stronger anharmonicity and larger phase space for scattering of softer TO modes and heat-carrying acoustic modes~\cite{delaire_2011,Shiga2012,Romero2015,Murphy2016}. This result together with our calculations of the electronic thermoelectric properties near the phase transition suggests that $ZT$ can be substantially increased (roughly doubled) by driving PbTe to the verge of the soft-mode phase transition due to external stress. This effect arises because soft TO modes that cause low $\kappa_L$ in PbTe are not the dominant e-ph scattering channel. Polar LO phonon scattering is the strongest scattering mechanism and is fairly constant towards the phase transition. This specific feature of the e-ph interaction in PbTe allows the soft-mode phase transition to effectively increase scattering of heat-carrying acoustic phonons while preserving the electronic thermoelectric properties, hence materializing the "phonon glass-electron crystal" idea without nanostructuring or alloying.}

\section{Conclusion}

Using first principles calculations, we investigate electronic thermoelectric transport in n-type PbTe driven closer to its intrinsic soft-mode phase transition via lattice expansion due to external stress.
We find that even though soft transverse optical modes interact more strongly with electrons electrons responsible for transport
when PbTe approaches the
phase transition, soft mode scattering is relatively weak
compared to longitudinal optical mode scattering in contrast to SrTiO$_3$~\cite{Zhou2018}.
The dominant electron-longitudinal optical phonon interaction and the electronic thermoelectric properties are
fairly insensitive
to the proximity to the phase transition.
\red{Consequently, the thermoelectric figure of merit $ZT$ of n-type PbTe can be enhanced by increasing the proximity to the soft-mode phase transition via lattice expansion since the lattice thermal conductivity can be sizeably reduced~\cite{Romero2015}.}
A few well-known high-performance thermoelectric materials, such as GeTe and SnTe, undergo the same type of the soft-mode phase transition with temperature.
%The strategy proposed in
Our work suggests new opportunities to further improve the $ZT$ values of these materials.

%We computationally demonstrate a new strategy to increase the thermoelectric performance of PbTe by driving the material closer to its intrinsic soft-mode phase transition, here simulated via lattice expansion due to external stress. Using first principles calculations, we show that the zone center transverse optical phonon modes substantially soften near the phase transition, while longitudinal optical and acoustic phonon modes are almost unchanged. Regarding electron-phonon interaction, we find that even though soft modes interact more strongly with electrons when PbTe approaches the phase transition, soft mode scattering remains relatively weak compared to longitudinal optical mode scattering. The dominant electron-longitudinal optical phonon interaction and the electronic thermoelectric properties are fairly insensitive to the proximity to the phase transition. As a result, the thermoelectric figure of merit $ZT$ in n-type PbTe can be considerably increased simply by increasing the proximity to the soft-mode phase transition via lattice expansion since the lattice thermal conductivity can be sizeably reduced~\cite{Murphy2016}. A few well-known high-performance thermoelectric materials, such as GeTe and SnTe, undergo the same type of the soft-mode phase transition with temperature. The strategy proposed in this work suggests new opportunities to further improve the $ZT$ values of these materials.

\section*{Acknowledgement}
%We thank Olle Hellman for kindly providing us the code that implements the TDEP method. 
This work is supported by Science Foundation Ireland under grant numbers 15/IA/3160 and 13/RC/2077 (the later grant is cofunded under the European Regional Development Fund), and Natural Science Foundation of Jiangsu Province under grant number BK20180456. We also acknowledge the Irish Centre for High-End Computing (ICHEC) for the provision of computational facilities.

\bibliography{biblio}

\end{document}